\documentclass[preprint]{aastex701}
\usepackage[flushleft]{threeparttable}
\usepackage{graphicx}

\newcommand\HI{H\,\textsc{i}}

\begin{document}

\markboth{Near Field}{Jacobs}

\title{Mapping Cosmological Signal Scales to Beam Calibration Requirements in 21cm Experiments and Implications for Near-Field Measurement}

\author{Daniel C. Jacobs}
\affiliation{School of Earth and Space Exploration, Arizona State University, Tempe, AZ, 85282}
\email{daniel.c.jacobs@asu.edu}

\begin{abstract}
Instruments targeting 21~cm emission at high redshifts need a spectral dynamic range of better than ten thousand to distinguish the 21~cm background against bright foregrounds.  Systematics  arising from the antenna pattern are a leading limitation for current instruments and must be addressed in future experiments. Antenna pattern measurements could help reach this precision.  Pattern measurements are complicated by the large scale of the instruments and interaction with the local environment. In-situ beam mapping methods have been investigated but the required accuracy remains ill defined. An aspect of beam measurement which must be defined is whether the calibration source is in the near or far field. Near field measurements require more elaborate measurement and such an expense must be well motivated. The far field distance is set by the effective size of the antenna. Reflections and interactions with surroundings extend the effective size of the antenna to scales well beyond the physical aperture. Here we give a new, instrument-agnostic method for calculating beam calibration requirements. Using theoretical 21cm models and instrument noise we prescribe bounds on the geometric reflection size scales. Systematics on these scales must be shown via measurement to be below the noise. This prescription depends weakly on instrument-specific noise and for interferometers, on the characteristic baseline length, but is otherwise independent of any detailed simulation of antenna or analysis pipeline.  
  Example calculations for HERA-like and EDGES-like instruments find theoretical cosmological structures map to reflection scales of 100~m.  Such antenna scales put the far field distance such that ground-based transmitters would need to be close to the horizon and drone sources well above typical or legal operating heights. A near-field measurement approach is necessary. Phase-locked systems have been demonstrated with promising results but more work is necessary to validate an antenna pattern at the necessary dynamic range.
\end{abstract}
\accepted{to the Journal of Astronomical Instrumentation}

\section{Introduction}

Instruments observing the redshifted 21 cm line at cosmological distances are limited by instrument precision. The antenna pattern is one of these unknown properties. Drone-based mapping could potentially help improve knowledge of the antenna beam but how accurate must drone measurements be? Which of the many developments in drone-based beam mapping are best suited to the problem? This paper is a step towards defining what is necessary by observing that theoretically predicted cosmology signals map to large physical scales. Using this result size in the far field results in a distance well-beyond drone flight levels, justifying the expense and difficulty of near field measurements. The effective length scales corresponding to the cosmology modes are good to avoid when choosing cable lengths, antenna spacings, or digital tolerances.

Measurement of the 21cm line from Hydrogen at cosmological distances 200My to 2Gy after the Big Bang traces the large scale properties of first stars and galaxies as well as the underlying physics driving early times such as massive neutrinos, supermassive black holes, and non-standard cosmology 
\citep{Morales:2010p8093, Furlanetto:2006p2267, Madau:1997p2232}. The predicted signal has a monopole or ``global'' component which evolves with redshift and on top of this is a fluctuating component which varies with angle and frequency\citep{2008PhRvD..78j3511P}.  At lower redshifts, 1 to 2,  baryon acoustic oscillations drive the galaxy power spectrum\citep{Chang:2008}. During reionization between redshifts 5 to 12 ionized regions make for strong fluctuations on the 10s of milliKelvins. Additional power is predicted at redshift 18 during the time of the first stars.

Interferometers targeting fluctuations include the MWA \citep{Wayth:2018,Beardsley:2019}, LOFAR \citep{Yatawatta:2013p9699}, HERA \citep{Deboer:2017PASP,Berkhout:2024}, SKA-low \citep{Mellema:2013p10035,Labate:2022}, CHIME \cite{Bandura:2014,Newburgh:2016}, and OVRO LWA \cite{Eastwood:2019}. Single antenna experiments have reported measurements of the global average. Evidence for an absorption trough at redshift 18 has been reported by the EDGES experiment \citet{Bowman:2018} while a followup by SARAS3 is in tension with that result\citep{Bevins:2022}.

The observed spectrum is the product of the chromatic antenna pattern with the spatially and spectrally varying foreground sky. 
The observation goal is to distinguish the background 21cm signal against foregrounds which are 10,000 times brighter\cite{2005ApJ...625..575S}. The 21cm signal varies both spectrally and spatially. The measurements reported by the interferometers are limited by systematics -poorly controlled aspects of the instrument or environment- rather than receiver noise \cite{Nunhokee:2025,HERACollaboration:2023, Mertens:2025}.  Global instruments are also very sensitive to instrument error, but with fewer data points it can be harder to distinguish from a possible detection\cite{Sims2020}.The instrument goal of clean separation of smooth spectral structure from spectral line modes is made difficult by instrumental imperfections which introduce unwanted chromatic response\cite{Thyagarajan:2016, Trott:2016,Trott:2017,LiuandShaw:2020}, and by any unmodeled spectral structure of foreground objects because best available foreground maps also have uncertainty\cite{Wilensky:2025}.

Systematics have proven difficult to remove because errors in instrument model used for calibration or subtraction result in excess power spectrum. See for example the most recent NenuFAR result \citep{Munshi:2024} where state of the art foreground modeling, calibration, and flagging decreases foreground residuals by three orders of magnitude, but the residuals are still 100 times what is expected for the cosmological signal. Meanwhile methods which build a model using the data itself run the risk of overfitting \citep{2013MNRAS.433..639P,Patil2016,Cheng:2018,KernandLiu:2021}. Filtering models that rely on the data may be shown to work safely in simulation, but results will remain freighted with caveats limiting interpretation \citep{Aguirre2021}. More accurate instrument models are needed. One of the most uncertain and difficult to control aspects of a low frequency telescope is the radiation pattern, or ``primary beam''.

This paper makes the following contributions to the literature. First, it introduces a framework that maps line-of-sight cosmological modes directly to the physical length scales of instrumental systematics. This framing provides a science-driven alternative to simulation-based requirement setting that is computationally lightweight and more transparent. More importantly it directly exposes how cosmological observables map to instrumental tolerances. Second, it derives bounds on allowable systematics on these scales using foreground and noise limits, with only limited dependence on instrument-specific details. Third, it demonstrates, using representative global signal and interferometric experiments, that these constraints place calibration sources firmly in the near field. Within this framework, the requirement for near-field beam measurement follows directly from cosmological detectability requirements.

\subsection{The Antenna Pattern}

An antenna's design and its surroundings introduce spectral distortions.
When an antenna is placed on a ground plane over soil, differences in conductivity can change the beam at a few percent level \cite{Spinelli2018}. Beam uncertainty mixes, through calibration or foreground removal, with uncertainty in foreground measurements to produce spectral structure \cite{Gehlot:2021,Ewall-Wice:2017}. A nearby reflective item in the environment can introduce spectral structure even at relatively long ranges  \cite{Rogers:2022}.   These studies used simulations and antenna reflectivity measurements to infer the response and do not confirm with direct measurement. However, in some cases, direct measurements using airborne transmitters have been used, for example to confirm the gross properties of SKA antennas predicted by simulation \cite{Acedo:2018,Virone:2021} and to confirm simulations indicating  unplanned chromaticity due to interactions within SKA stations \cite{paonessa:2023}. 

\subsection{Antenna Pattern in Interferometers}
In an interferometer, the antenna pattern is further complicated by other antennas nearby. This pattern is mixed into the point spread function (PSF) due to the antenna layout. Limited sampling in the $uv$ plane causes the PSF to evolve with frequency resulting in the so-called foreground wedge in the 2D power spectrum. Further distortions due to mutual coupling, cable reflections or similar effects place copies sky at other locations in the power spectrum space. This is described in more detail later in section \ref{sec:pspec-how-it-works}. 

 The reduction of spectral structure in the primary beam has been considered in the design of upcoming arrays.  
 Coupling between antennas or other surroundings is a strong source of spectral structure which, in simulation, has generated systematic floors in the SKA for both power spectrum \cite{Ohara:2025} and a phased-array global signal \cite{Stavely-Smith:2025}. Those studies considered coupling internal to the phased array antenna station which results in spectral structure in the beam. However study of HERA data and simulations have shown that coupling \emph{between} antennas results in high delay systematics which appear at all observed $k$ modes \cite{Rath:2024}.

\subsection{Breaking the scale ambiguity in beam calibration}

A central challenge in beam calibration for 21\,cm experiments is determining what physical aspects must be controlled and measured to sufficient accuracy. While simulation studies have shown that beam errors at the level of 1\% in the main lobe and 10\% in the side-lobes can be sufficient to avoid contaminating the 21~cm signal \citep{Ewall-Wice:2020}, these requirements do not directly specify what measurements are needed in practice, nor which physical effects dominate the error budget. More importantly, it is not clear how to generalize such requirements across different instruments and observing strategies.

This leads to a circular problem. The dominant sources of beam error must be identified through measurement, but the interpretation of those measurements depends on whether they are made in the near or far field. If large-scale effects dominate and measurements are interpreted using far-field assumptions when in fact they are in the near field, the resulting errors may be uninterpretable. Conversely, designing a measurement campaign to operate in the near field without knowing whether such scales are relevant may impose unnecessary complexity and cost.

Simulations of specific instruments \citep[e.g.][]{Ohara:2025} provide valuable insight into the underlying physical effects, but they rely on detailed models which will not reflect the as-built design. What such studies have identified is the significance of large scale phenomena such as mutual coupling, but without an exhaustive consideration of all possible effects are not able to conclusively put an upper limit on the relevant instrument size scale.

The distinction between near- and far-field measurement depends on a characteristic size scale $D$, since the far-field distance $R$ is given by
\begin{equation}
    R \gtrsim \frac{2D^2}{\lambda}.
\end{equation}
In standard antenna theory, $D$ is taken to be the size of the aperture \citep{balanis}. However, for 21\,cm instruments, the relevant scale is not necessarily the antenna itself, but the size of whatever physical structure introduces chromatic systematics.

If beam errors are dominated by manufacturing variations, each antenna can be treated in isolation and $D$ is set by the antenna size, placing typical drone measurements comfortably in the far field. In contrast, if distortions arise from reflections or coupling involving the surrounding environment, $D$ may correspond to distances of tens to hundreds of meters, placing realistic measurements firmly in the near field.

To resolve this ambiguity, we introduce a science-driven approach that sets the relevant scale $D$ using a cosmological signal model. By identifying the spectral modes that must be measured to detect the 21\,cm signal above foregrounds and noise, we map those modes to physical length scales in the instrument. This breaks the circularity by determining, prior to measurement, which spatial scales must be controlled and therefore whether near-field measurements are required.

First we will explore how the cosmological scales of interest map to physical scales starting with a short review of 21cm cosmology methods (S\ref{sec:how_it_works} to explain how beam uncertainty causes systematic errors (S\ref{sec:beam_uncertainty}) and then use cosmology simulations to define the range of modes which are not occluded by noise or foregrounds (S\ref{sec:cosmology_mode_scales}.  From there we can map these modes to far field distances which are well out of reach of standard drone systems. Having established the requirement for drone mapping to be in the near field we review previous work towards this goal (S\ref{sec:drones_and_mapping}) and close with a brief proposal (S\ref{sec:drone_prescription}) for a system solution which combines several ideas.

\section{Deriving Beam Mapping Requirements from cosmology models}

\subsection{From 21cm model to instrument spatial scales}
\label{sec:how_it_works}
Let us now quantify how the cosmological scales of interest map to instrumental scales, beginning with a short review of the theoretical 21~cm signal, and how we might expect it to be structured in redshift/frequency.

Between the end of reionization at redshift 5 (200MHz) up to the first appearance of stellar Xrays at redshift 20 (57\,MHz), intergalactic Hydrogen is visible as a spectral distortion on top of the Cosmic Microwave Background.  At redshifts below 5 (1420 to 200MHz), neutral \HI in galaxies trace large scale structure. Amplitudes range from -200 mK (absorption) to tens of mK in emission, depending on redshift/frequency. Models of the signal evolution predict structure distributed over scales as small as 10 arcminutes and 100kHz up to several degrees and several MHz. The fluctuation power spectrum is expected to be nearly isotropic and drop as a power law with increasing $k$.  Power spectrum measurements expect to see fluctuations between 10 and 100mK. A fiducial simulation, made using ZEUS\footnote{Zippy Early Universe Solver for 21cm \citet{Munoz2023}, \url{https://github.com/JulianBMunoz/Zeus21}}, is shown in Figure \ref{fig:gs_pspec_compare}. Two representative parameter choices are shown to illustrate a typical range of theoretical possibilities.  

\subsubsection{Global Measurements with a Single Antenna}
\begin{figure}[htb]
    \centering
    \includegraphics[width=0.45\columnwidth]{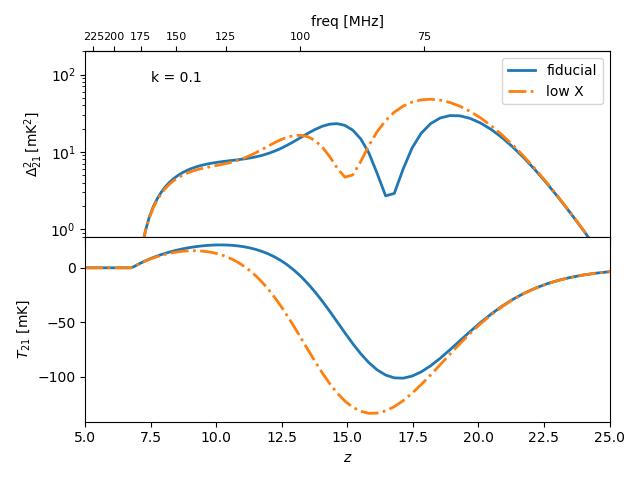}
    \includegraphics[width=0.45\columnwidth]{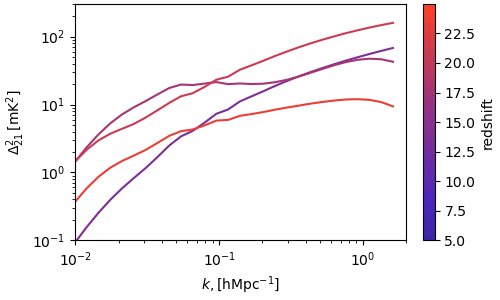}
    \caption{A representative selection of 21cm models generated using ZEUS \cite{Munoz2023}. These models illustrate a typical range of spectral scales predicted by theory. The left two-panel plot illustrates how the global signal (lower) and power spectrum (upper) evolve with frequency. Relatively small changes in one parameter like the XRay luminosity of early objects can cause large changes. The power spectrum (bottom plot) shows how the power spectrum (the fiducial model is shown here) changes with redshift as structure evolves. The $k$ axis tracks scales at equal parts along three dimensions: angular size and frequency, which maps to distance. Figure \ref{fig:Pk_vs_dly} shows these in instrumental units of power vs spectral mode/delay.}
    \label{fig:gs_pspec_compare}
\end{figure}

The global signal, illustrated in Figure \ref{fig:gs_pspec_compare}, is predicted to have a spectrum with absorption and emission features which track the dominant physics of the proto-intergalactic medium. Any energy inputs or decrements occurring in the dense early universe will have an impact on the global spectrum. The largest feature is an order-100mK deep, 50MHz-wide absorption trough expected somewhere  driven by a sequence of UV and Xray emission from the first stars. At higher frequencies an excess due to emission by hot neutral \HI is predicted to extend to 150MHz or higher. 

\subsubsection{Power Spectrum}
\label{sec:pspec-how-it-works}
\label{sec:power_spectrum_scales}
Deviations around the mean can be measured with an interferometer.  The observed frequency maps back to the implied redshift of the 21cm line. For small frequency changes the frequency corresponds to line of sight distance, while larger frequency separations see causally distinct epochs. In the absence of noise, an imaging spectrometer could measure a 3D image cube at each redshift, sampling back along our lightcone. In practice, sensitivity is a limiting factor so instruments instead measure the power spectrum. The power spectrum found by extracting co-evolutionary chunks of the spectrum, forming a 3D Fourier transform of the observed volume and averaged spherically across all three dimensions. For a complete explanation of this geometry, see \citet{LiuandShaw:2020}, their section 3.3 and Figure 5.

The simulated power spectra shown in Figure \ref{fig:gs_pspec_compare} illustrate how a typical model evolves with redshift --the model is described in detail in Section \ref{sec:power_spectrum_scales}. It also shows how the power spectrum evolution varies under reasonable changes to astrophysical inputs. For example, if the first star XRay luminosity is lower by a factor of 2, the power spectrum rises to a higher peak and passes through a minimum later.

\subsubsection{Systematics in the power spectrum}
\begin{figure*}[htb]
    \centering
    \includegraphics[width=\textwidth]{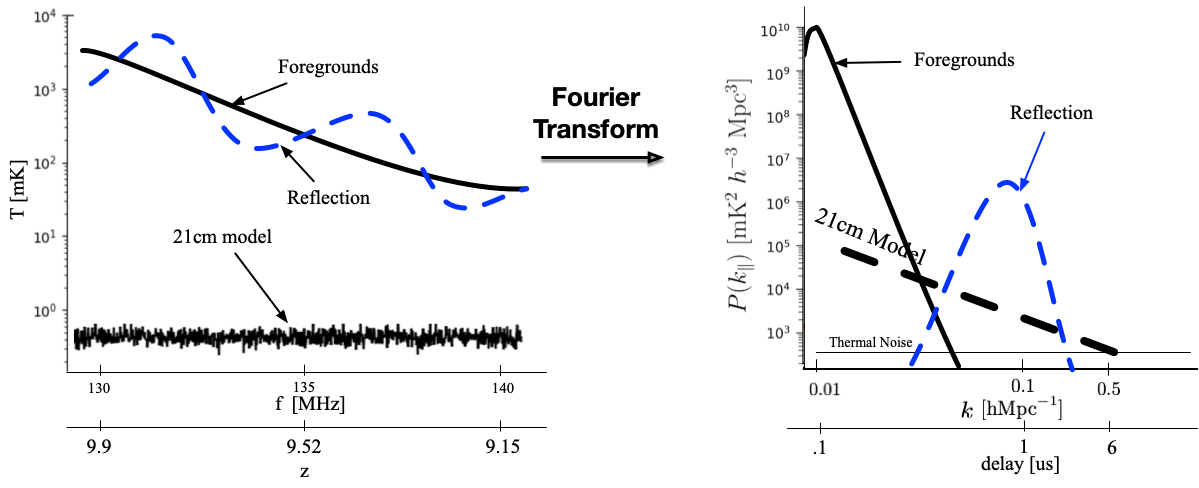}
    \caption{Illustration of internal reflection in a power spectrum along a single line of sight vs frequency/distance (left)   and Fourier domain power spectrum (right) in $Pk$ density units.  The smooth spectrum foreground is separable from the background 21cm signal in the Fourier domain, but the periodic gain structure of a reflection couples power to normally uncontaminated Fourier modes. The high level instrumental requirement is therefore that all modes due to reflections be lower than thermal noise. This holds only in the delay range where signals are forecast to dominate over intrinsic foregrounds and noise.}
    \label{fig:foreground-reflection-cartoon-pspec}
\end{figure*}

The Fourier transform across frequency motivates attention to resonant structures in the instrument.
Foreground objects at these frequencies are primarily smooth spectrum synchrotron or free-free emission (see eg \cite{Ross2024}). These will have an intrinsic power spectrum dominated by the lowest $k$ modes, separating the power from the (theoretically) flat power spectrum of the 21cm signal. This is illustrated in Fig \ref{fig:foreground-reflection-cartoon-pspec}.

\begin{figure}[htb]
    \centering
    \includegraphics[width=0.5\linewidth]{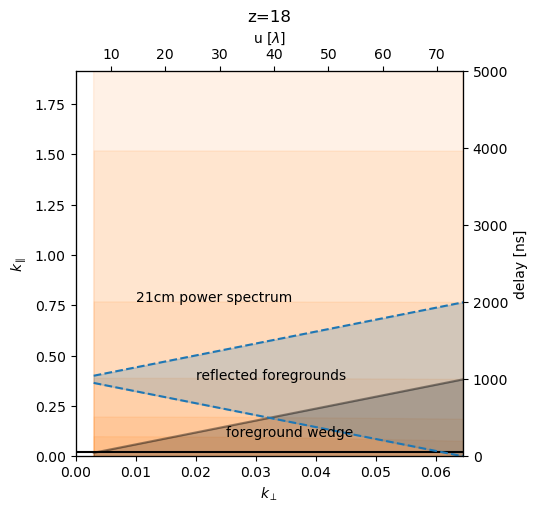}
    \caption{Schematic illustration of the 2D power spectrum wedge for a fiducial observation of 10MHz at redshift 20 obtained by gridding visibilities to a uvf cube, Fourier transforming along frequency to get delays and then averaging cylindrically in u. Smooth spectrum foregrounds would lie below the black line corresponding to 1/bandwidth but the PSF scatters power the length of the baseline in blue. Internal reflections place copies of the sky at the reflection delay. Here a simple reflection of 300m is indicated in dashed lines. }
    \label{fig:wedge}
\end{figure}

A useful stop between raw 3D measurements and a 1D power spectrum is the 2D cylindrical average over angle on the sky. This shown schematically in \ref{fig:wedge}.

Each baseline of the interferometer observes a smooth spectrum point source as an oscillatory function in frequency with a period set by the baseline length and elevation angle. Longer baselines (ie higher $k_\perp$) have power at higher delay (ie $k_\parallel$). When nearby baselines are gridded in the $uv$ plane and averaged together the effect is reduced roughly by the degree to the accuracy of sampling in the $uv$ space. The resulting 2D power spectrum in $k_\perp$,$k_\parallel$ makes a wedge shape \cite{Parsons:2012,Liu2014b,Thyagarajan:2015,Morales2018,Mertens:2025}.

Figure \ref{fig:wedge} shows some of the relevant landmarks of the 2D power spectrum space. Sources at the horizon have the highest delay and appear along the diagonal line; sources towards the field center are below the line, filling the body of the wedge. In this 2D space, the predicted power spectrum is roughly constant with angle, dropping strongly with $|k|$ as illustrated by the orange contours.   Though attenuated by the beam pattern, it is the sources towards the horizon which push closest into the otherwise uncontaminated $k$-space and encroach on the brightest power spectrum modes.

Given the known foreground levels and best achievable noise, both of which are much larger than the cosmological signal, only a subset of spectral modes are predicted to have detectable signal to noise. The upper edge of the wedge is a practical boundary on the longest delay scales that an instrument must measure accurately. In Section \ref{sec:power_spectrum_scales} we will use this fact  to bound the relevant range of power spectrum modes. 

\subsubsection{Spectral Modes}

Any unintended instrumental spectral modes cause power to be moved to higher $k_\parallel$ modes which are nominally clean. A spectral mode corresponds to a physical length scale. A spectral ripple of period $df$ corresponds to a length scale $c/df$, up to factors of order unity depending on path geometry. For example, a reflection off a nearby object at distance $D$ introduces a ripple with a frequency period $c/2D$.\footnote{Periodic spectral features can also be caused by lumped element circuits or reflections within the signal chain cabling. In this paper we are aiming to understand the impact of physically large antennas or arrays and will leave other sources of aside for now.}.

\subsection{How beam uncertainty causes systematic errors in 21cm measurements}
\label{sec:beam_uncertainty}
The antenna beam can introduce uncertainty in many ways. The problem has been studied for both single-antenna and interferometer experiments. Though those looking at each instrument type have sometimes adopted  different language and approaches, the fundamental concepts are the same: unexpected behavior mixes bright foreground into cosmological modes. 

\subsubsection{Interferometer Beam systematics}
Interferometer simulations have found that bright sources far from the pointing center, though attenuated significantly by the beam, enter in on modes set by the length of the baseline \citep{Thyagarajan:2013p10039, Thyagarajan:2015a, Mort:2017}. This is the much-discussed \emph{wedge} \citep{2015:ThyagarajanConfirmationwidefield,Pober:2016ApJ...819....8P, LiuandShaw:2020}.

These sources can be subtracted, but only to the accuracy of the instrument model. Consider the 750Jy source Fornax A, one of the brightest sources in the southern sky. During certain times of night it climbs well into the primary lobe of both HERA and MWA with amplitudes ranging from 0.1 to 0.7 relative to the peak depending on frequency. When Fornax A is as high as it gets in the HERA beam it rises to about the half power point. A 0.1\% beam error in subtraction would leave a residual source flux of 750mJy. Now suppose the gain error is made in a gaussian random way on each channel. This amounts to a noise-like variance with a flat power spectrum amplitude of about 3x10$^5$mK$^2$ in $\Delta^2$ units, much brighter than the forecast spectrum shown in Fig. \ref{fig:ff_pspec} which is never higher 200mK$^2$.

A 0.1\% beam error causes a systematic which is 1000x brighter than the expected background signal shown! In practice, the variation is  smooth with frequency leading to a spectrum that looks more like the cartoon illustration in Figure \ref{fig:foreground-reflection-cartoon-pspec}.
 
 More detailed electromagnetic models do improve the overall accuracy of flux reconstruction (eg \citep{Sutinjo:2015}) but electromagnetic models  cannot account for as-built variations between elements within an array, which has been shown to be significant and largely unavoidable without large added expense \citep{Neben:2016a}.   Low-frequency arrays are particularly susceptible due to use of materials such as wire mesh, metal channel, and/or plastics, antenna to antenna assembly variance, corrosion, plant incursion, rain, and soil variability.  Antennas require in-situ characterization.

An inaccurate beam model affects calibration accuracy. Variation in signal chain performance causes each antenna to have an unknown gain and phase which must be found experimentally. The usual method is to use astronomical calibration sources, limiting the model to sources currently located in well-known parts of the beam. However,  incomplete foreground models have been shown to cause leakage similar to reflections\citep{Barry:2016}. Mismodeling sources that are nevertheless present in the data causes systematic errors.  

Calibration error is not just limited to sky modeling.  Arrays like MWA Phase II and HERA are using baseline redundancy to calibrate using linear algebra \citep{Liu:2010p10391,Zheng:2014p10467,Ali:2015}. Though this method minimizes reliance on the sky model it does assume the beam is the same from antenna to antenna. Additional investigation is necessary to understand the impact of beam non-redundancy, but it is already clear that these deviations introduce chromatic systematics when calibrated against the sky \cite{Li:2019, Barry:2016}.

Another measurement challenge is the wide spectral bandwidth needed to capture the spectral evolution shown in Fig \ref{fig:gs_pspec_compare}. 
Instruments like HERA, MWA, and EDGES span fractional bandwidths of two to three, from $\sim$100 to 200 MHz, or 50 to 100MHz depending on the cosmological redshift range of interest. 
Spectral variation of the beam has been found to cause otherwise smooth foregrounds to contribute more complicated contaminating structures \citep{Thyagarajan:2016, Ewall-Wice:2016b}.  In early observation simulations, focus centered on the amplitude of the beam near the horizon \citep{Thyagarajan:2015a, 2015:ThyagarajanConfirmationwidefield} and the variation of the beam from antenna to antenna \citep{Neben:2016a}. However first results from HERA, which uses large, densely packed dishes has turned attention to mutual-coupling between antennas \citep{Kern:2019,Fagnoni:2021,Josaitis:2022,Rath:2024}. 

Mutual-coupling can be understood as a passive gain change introduced by neighboring elements but much of this can be modeled as alternate signal paths within or between antenna elements. Multi-path effects have been predicted between HERA feeds using an electromagnetic model\cite{Fagnoni:2021}). Due to computational limitations only a subset of the array was simulated. Using this model as a reference a semi-analytic model assuming a re-radiation mechanism was used to extrapolate to the complete HERA array and generate simulated data where it broadly shares similar features \cite{Josaitis:2022}. The smallest reflection scale is the 14.25m distance between antennas and the largest relevant distance is the 300m diameter of the core.

Effects due to surroundings have been identified as a candidate for gain and phase uncertainty in the MWA antennas as well.  \citet{Chokshi:2021} measured the beam patterns of MWA tiles using satellite transmissions finding beam pattern deviations of more than 5dB in the primary and secondary sidelobes.  These are noted to be large enough to have a significant impact on science cases like the 21cm program. Reasons for deviation include dead dipoles in each tile, but also surrounding vegetation and rocks.  Testing of these hypotheses was left for future work and no distances to potentially reflecting bushes were given. A reasonable scale to consider is the size of the Phase II hexes which are just under 100m across (see Figure \ref{fig:telescope_aerial})

\subsubsection{Global experiment beam systematics}
The antenna beam is also a dominant source of uncertainty in global experiments for much the same reasons. A total power instrument measures the integrated product of the sky and the beam to produces a single spectrum which samples different sky regions as the Earth rotates.  One analysis path is to normalize out the expected foreground spectral shape \citep{Bowman:2018}. However, the beam models show signs of being fragile, with variations between solvers that could be large enough to introduce systematics at a relevant level\cite{Mahesh:2021}. 

Given the modeling uncertainty, global experiments have designed antenna configurations that minimize undesired modes or somehow average them out. 

Where interferometers are unavoidably electrically large, global signal experiments are not intrinsically so. But in practice, the presence of surroundings introduces structure.

For example, \citet{Bradley:2019} suggested that the EDGES absorption signal could be a combination of loose ground plane connections and a layer of moisture leading to a standing wave. Effects like this have driven single antenna experiments to add more control over their environments with larger mesh ground planes \citep{Bowman:2018}, on a lake\citep{Nambissan:2021}, or averaging over different antennas\cite{Saxena2023}.  As an example of large  size scales note the 50m diameter of the EDGES3 ground plane and the distance to the nearby electronics hut (300m away, see Figure \ref{fig:telescope_aerial}) which has been covered with absorptive tile.

 Reflections in the environment or from mutually coupled antennas all emerge as spectral and spatial ripples at a small percentage of the mean gain. A 0.1\%  spectral ripple is sufficient to be a leading systematic if the ripples occur at the scales of the cosmological background modes\citep{Lanman:2019}.   Relevant sizes for a selection of instruments are listed in Table \ref{tab:antenna_scales}.  Later in section \ref{sec:farfield} we discuss how these scales are all in the near field for drones. But first, let us compare the spectral scales of these instruments with the size scales that matter for the 21\,cm signal.

\begin{table}[]
    \centering
    \begin{tabular}{c|c|c|r}
    \textbf{Instrument} & \textbf{Size} & \textbf{$c/$Size} &   Dimension\\
    \hline
       HERA  & 300m & 1MHz & size of core \\
       MWA  &  70m & 4.3MHz & hex size \\
       EDGES & 50m & 6MHz& size of ground plane\\
    \end{tabular}
    \caption{Relevant physical sizes for some 21\,cm arrays.}
    \label{tab:antenna_scales}
\end{table}

\subsection{Cosmological Spectral Modes in Real Instruments}
\label{sec:cosmology_mode_scales}
What are the instrumental length scales that matter for the 21\,cm signal? Let us return to the ZEUS model introduced above and identify the points which can be detected above the noise and outside foreground contamination. This will set the range of modes that must be shown to be free from beam errors.   

To get a sense of sensitivity to astrophysics we calculated a reference simulation where the cosmological and various galaxy evolution parameters are kept at fiducial defaults and a second where XRay luminosity of star formation is reduced by a factor of 3 (see Figure \ref{fig:gs_pspec_compare}). Both models have a fairly smooth global signal response and similar power spectra with the strongest temporal evolution of the power spectrum on small k/large scales. Lower XRay input increases the depth of the absorption trough by almost 40\% and the peak power spectrum amplitude by a factor of two.

\subsubsection{Global Spectral Modes}
\begin{figure}[htb]
    \centering
    \includegraphics[width=0.45\columnwidth]{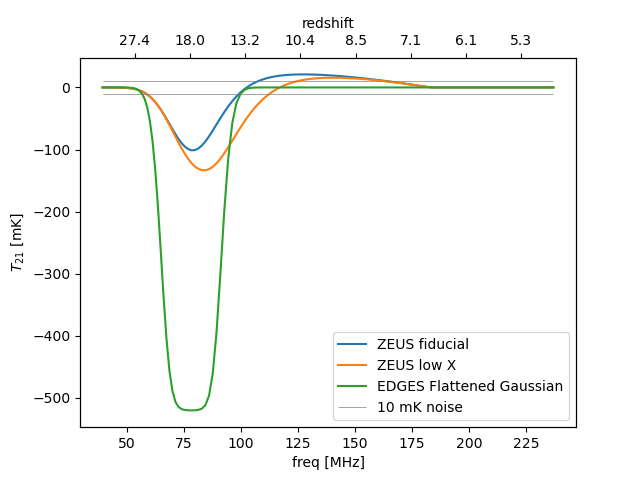}
    \includegraphics[width=0.45\columnwidth]{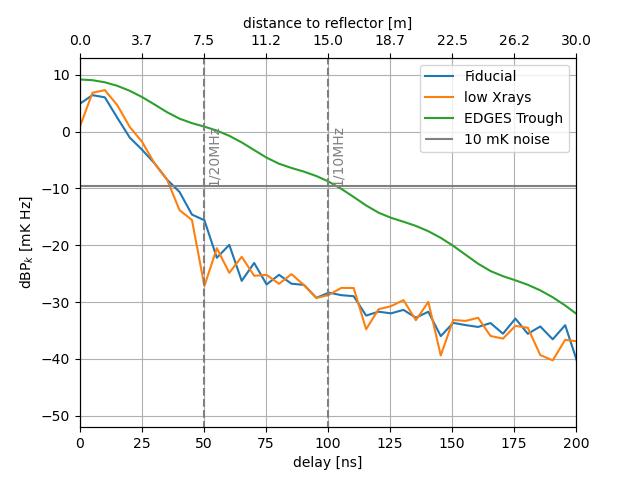}
    \caption{Fiducial global signals and their fourier modes. Theoretically-motivated global signal models are smooth, with little power at large delays. However, the flattened Gaussian proposed to explain excess observed by EDGES extends to large delays  corresponding to physically larger scales in the instrument.In practice even the largest scales are limited by noise, hrer}
    \label{fig:gs_delay_compare}
\end{figure}

\begin{figure}[ht]
\centering
    \includegraphics[width=0.5\columnwidth]{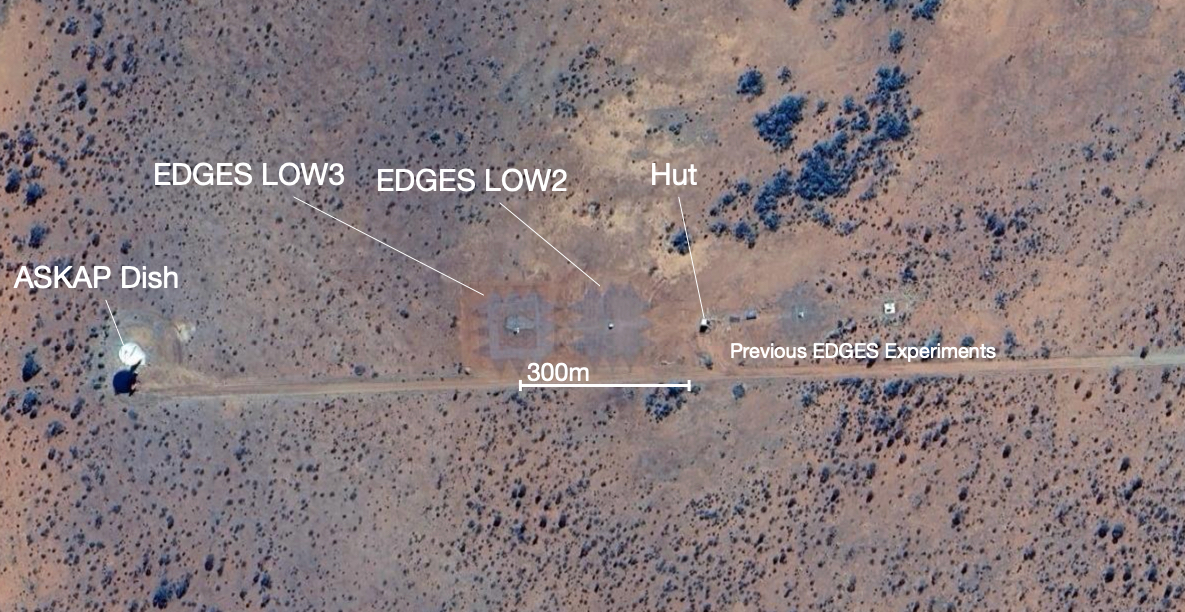}
    \includegraphics[width=0.5\columnwidth]{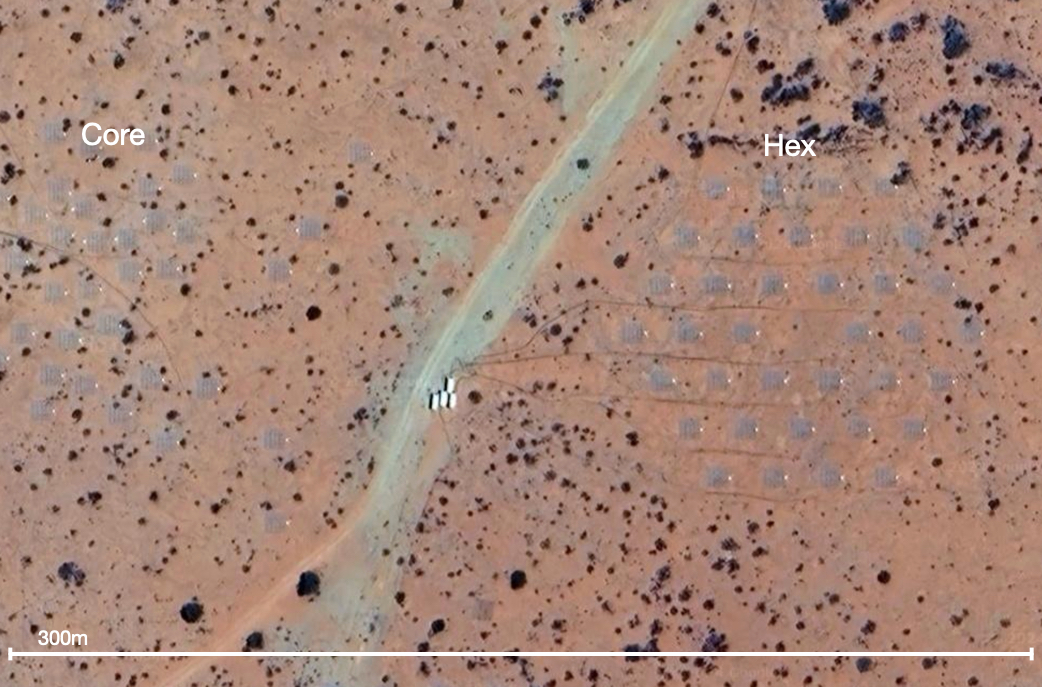}
    \includegraphics[width=0.5\columnwidth]{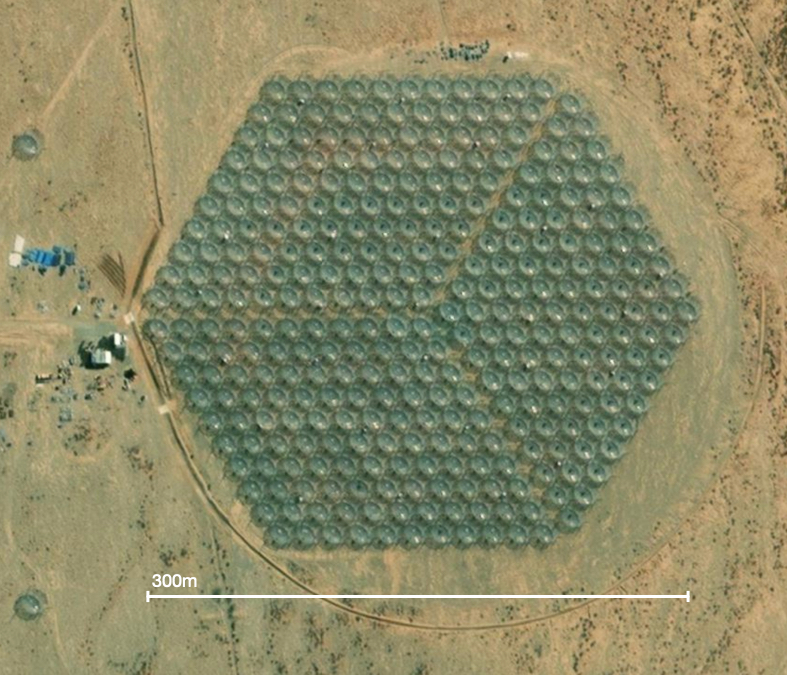}
    \caption{Aerial views of (top to bottom) EDGES, MWA, and HERA with a scale marker indicating 300m. Typical length scales are 300m and smaller. At redshift 20 a 300m reflection corresponds to a $k_\parallel$  of 0.0055 h/Mpc (251Mpc/h) a size of interest in many models and to 1MHz, roughly the scale at which the EDGES trough curves as it drops. }
    \label{fig:telescope_aerial}
\end{figure}

ZEUS predicts a global signal that evolves relatively slowly with frequency. Comparing the global signals in spectral Fourier space (Fig \ref{fig:gs_delay_compare}) one can see that both models have most of their power on delays slower than 1/20MHz with small differences between the two models.

Though the models are predicted to be smooth in frequency, observers have encountered faster evolution with redshift. In 2018 EDGES reported a trough  best modeled by a flattened Gaussian\cite{Bowman:2018}. The best fit descended 500mK in only 4 or 5 MHz and as we can see from Fig \ref{fig:gs_delay_compare} extends much further in delay space.

Though the reference simulations are smooth on the 10MHz scale, to properly eliminate uncertainty, a global signal instrument must resolve all doubts about ripple on scales at least up to 1MHz if not faster.

\subsubsection{Power Spectrum Scales}
\label{fig:power_spectrum_modes}
The spectrum modes relevant to a power spectrum measurement can be inferred from the reference model but there are also boundaries imposed by foregrounds and noise. Smooth spectrum foregrounds obscure the largest spectral modes (smallest $k$s) and noise dominates at the smallest scales (again, see \cite{LiuandShaw:2020} for more on this).  Both of these factors are somewhat instrument dependent. Here we demonstrate with a HERA-like instrument making a single high sensitivity measurement of a 28~m baseline measuring a single $k_\perp$ mode with 14m dish antennas from 60MHz to 200MHz. 

\begin{figure}[htb]
    \centering
    \includegraphics[width=0.5\columnwidth]{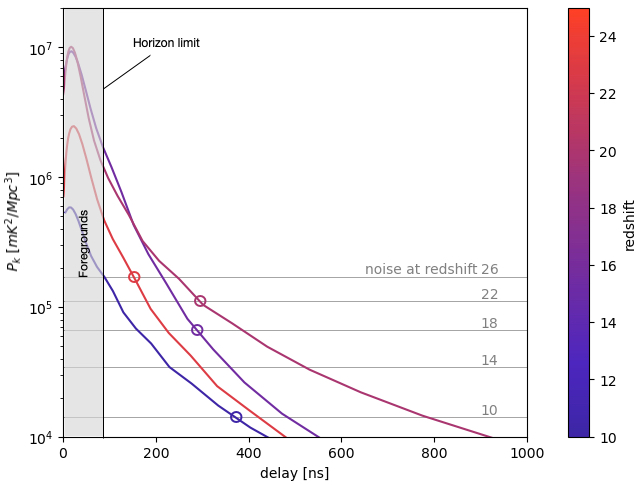}
    \caption{The range of detectable power spectrum modes is bracketed by foregrounds (black) and noise (grey). Here we illustrate this fact with the fiducial 21~cm ZEUS model shown for a range of redshifts and plotted in power spectrum density for a single $k_\perp$ mode.  Foreground and noise bounds chosen to approximate a deep HERA integration on a sensitive baseline (28m). Points where the power spectrum drop below the noise are circled. These set the upper bound on reflection ranges shown in Figures \ref{fig:pspec_window_in_delay}}
    \label{fig:Pk_vs_dly}
\end{figure}

First, take the theory power spectra in Figure \ref{fig:gs_pspec_compare} and scale into observer units, shown in Figure \ref{fig:Pk_vs_dly}\footnote{The scaling is approximated by converting to P$_k = \Delta^2/k^{-3}$ and estimating what a single baseline might look like by calculating $k_\parallel$ assuming a 28m long baseline. This approximation is valid where $k_\perp \ll |k|$. Which is true in this paper where $k_\perp$ is never more than 1\% of $k$.}. The $k_\parallel$ axis is converted to observer's nanoseconds and approximate foreground and noise regions are indicated.   In these units the power spectrum has the largest power at shortest delays dropping quickly towards larger delays and evolving with redshift/frequency.   Foregrounds, arising from power law processes like synchrotron, are smooth with frequency and are generally constrained to delays shorter than the length of the baseline.\footnote{This is also illustrated in Fig \ref{fig:foreground-reflection-cartoon-pspec}}  Gaussian noise is flat with frequency and therefore  also in the delay spectrum. Sky noise temperature rises with wavelength/redshift as a 2.26 order power law\cite{Mozdzen:2017}. The reference noise level shown is for a two-year HERA integration most recently estimated for a realistic HERA season by \citet{Breitman:2024}.  

\begin{figure}[htb]
    \centering
    \includegraphics[width=0.5\columnwidth]{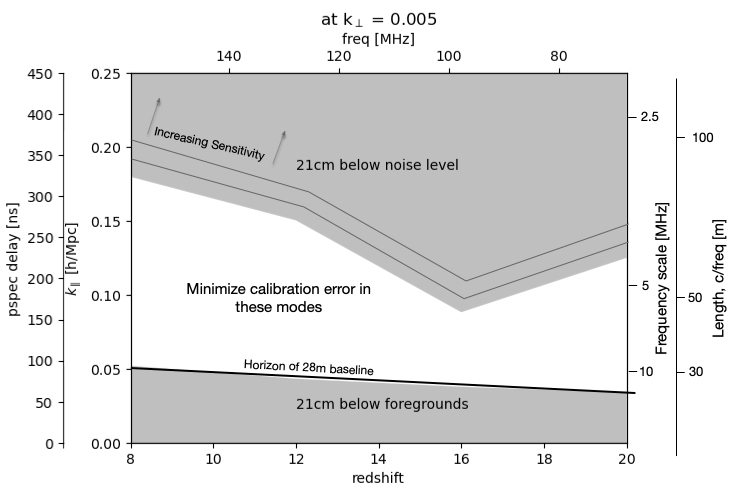}
    \caption{The useful range of delays where a fiducial 21\,cm cosmological model could be detected by a fiducial instrument as illustrated in Figure \ref{fig:pspec_window_in_delay}. Foregrounds set the lower bound and noise sets the upper. A 21\,cm interferometer needs to be very well calibrated on spectral scales between 2 and 10\,MHz}
    \label{fig:pspec_window_in_delay}
\end{figure}

The measurable modes occur between the foreground horizon and the point where the power spectrum drops below the noise level. In Figure \ref{fig:Pk_vs_dly} these points are marked with circles. This ``delay window'' is summarized in Figure \ref{fig:pspec_window_in_delay}. It is this range of delays where all unknown systematics must be below the thermal noise. The window gets wider as sensitivity increases and narrower as baselines get longer. It can also get get wider if foregrounds can be accurately subtracted or filtered opening up modes where the signal is supposed to be the brightest.   Of course this is all dependent on the shape of the power spectrum and its timing with redshift. If the spectrum were to have more power at higher $k$, or the higher power part of the spectrum evolution were to occur at later times, the maximum delay/frequency mode could increase. The variation with frequency gives some idea of the range of modes which should be measured.

Given a reasonable 21\,cm model and debatably generic instrument parameters one can say that the most important instrumental spectral modes, where the instrument response needs to be carefully calibrated, possibly with external sources, are between 2 and 10MHz or delays of 100 to 500 ns.

\begin{figure}[htb]
    \centering
    \includegraphics[width=0.5\columnwidth]{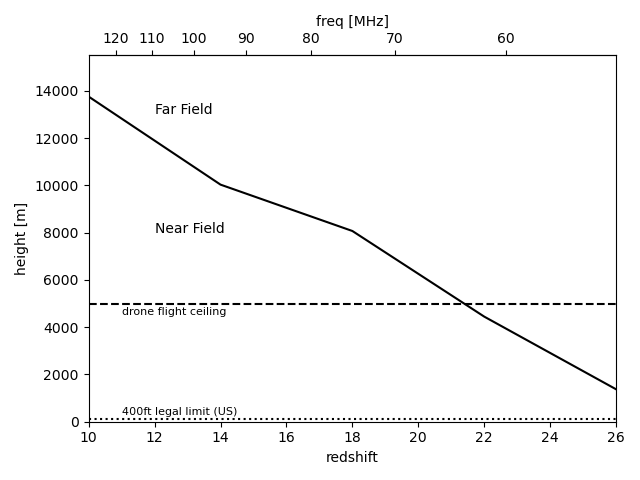}
    \caption{Translating the highest cosmological delay modes (from  Fig \ref{fig:gs_delay_compare}) into length scales and from there to the corresponding far field. Technically, the best commercially available drones have a flight ceiling of 5000 meters, but in practice legal and practical considerations put a limit of a few hundred meters. In the US the legal limit is 121m (400ft). Also note that ground based transmitters under the horizon are also largely within the near field.  }
    \label{fig:ff_pspec}
\end{figure}

\begin{figure}[htp]
    \centering
    \includegraphics[width=0.5\columnwidth]{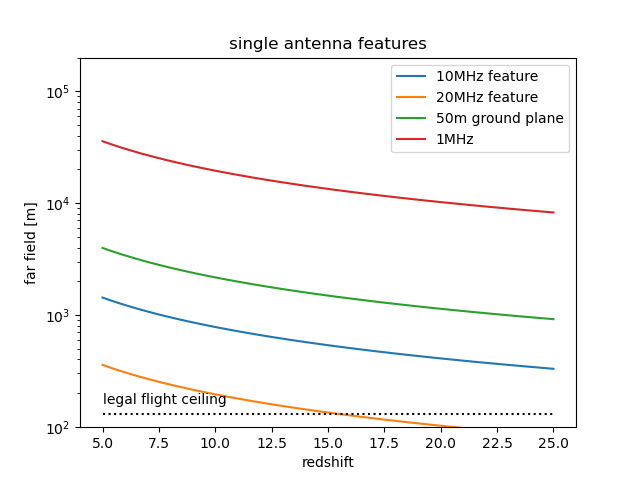}
    \caption{Far field distances for scales relevant to global experiments. Single antenna experiments have encountered  systematics with residual ripples having periods of 10MHz but sharper edged features as fast as 1\,MHz have been seen. Aiming to reduce the potential for resonances, ground planes have been increased, for example in EDGES3 where the ground plane is now 50m across.}
    \label{fig:ff_single_ant}
\end{figure}

\begin{figure}[htb]
    \centering
    \includegraphics[width=0.7\columnwidth]{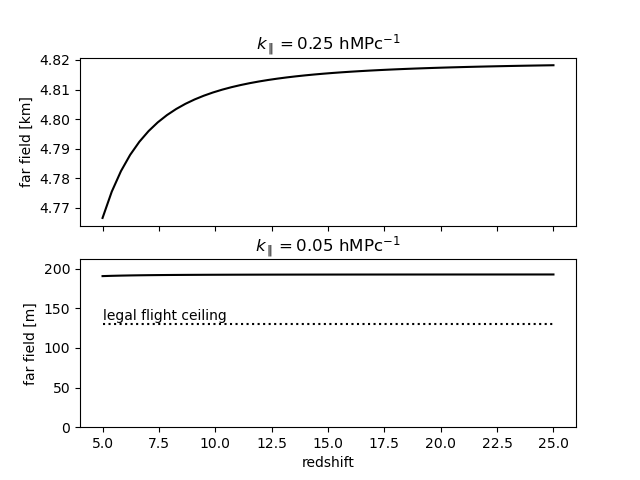}
    \caption{A reference comparison showing how $k_\parallel$ modes map to the far field. The legal flight limit in the US is 121m. }
    \label{fig:ff-vs-redshift}
\end{figure}

\subsubsection{Cosmology Requirements Imposed on 21\,cm Experiments}
In summary an admittedly brief inspection of a representative 21cm model suggests that, across the redshift range 5 to 25 the fastest spectral scale that matters is, adding in some margin and choosing a round number, around 1\,MHz or 100ns and the slowest 500ns.   This requirement roughly holds for both global and interferometric instruments.

\begin{table*}[htb]
\centering
\caption{Key quantitative results for example instruments HERA and EDGES}
\label{tab:placeholder}
\begin{tabular}{c c c c l}
\hline
 & Required Spectral Control (MHz) & Path lengths (m) & Implied far field (km) & Basis of estimate \\
\hline
EDGES & 1 MHz & 300 m & 7.5--30 km & \parbox[t]{4cm}{slope of observed trough} \\
HERA  & 3 MHz & 100 m & 0.9--3.6 km & \parbox[t]{4cm}{largest $k_\parallel$ mode predicted to be detectable} \\
\hline
\end{tabular}
\end{table*}

 \subsection{Drones and Fixed transmitters are in the Near Field}
 \label{sec:farfield}
 We have shown how the expected sizes of cosmological spectral modes drive instruments towards controlling precision at very long spectral scales. But how do these scales map to the instrumental problems?  
 
  A reflection of length D produces an interference pattern with a spectral ripple period $df = c/D$. Where D is the total length of an undesired path which radiation can follow in the instrument, for example, a reflection. \footnote{In practice the E-field does not follow simple geometric paths, particularly at sub wavelength scales or where materials form a complex apparent index of refraction. However since the power spectrum analysis breaks down the spectrum into spectral Fourier modes, we can arbitrarily re-scale from delay to length scale without loss of generality. Re-casting delay modes into reflection scales is a common approach, used for example to define the antenna requirements for HERA \cite{Ewall-Wice:2016b}.} A 1MHz ripple corresponds to a 300m path length or a physical distance of 150m.
 
 This size scale presents a challenge for drone-based measurements in the cosmic dawn and reionization bands where operational altitudes are in the near field. 

Instruments observing the Cosmic Dawn, Reionization bands between 50 and 200MHz have far field distances 
ranging from 30 to 150km. Far field limits for global experiments are shown in Fig \ref{fig:ff_single_ant} and interferometers  in Fig. \ref{fig:ff_pspec}.  

As a supplement to these projections one can also directly map $k_\parallel$ to delay scales and from there calculate the far field.  This is a useful mapping because it is independent of antenna type and presents a requirement defined by the spectral shape of the background signal.  For interferometers probing fluctuations it is set by the size of cosmological modes. The far field distance corresponding to two representative modes is shown in \ref{fig:ff-vs-redshift}. A $k_\parallel$ mode of $0.25hMpc^{-1}$, a reasonable fiducial for the largest measurable $k$ mode, maps to a far field of 76\,km with a weak dependence on redshift/frequency shown in Fig \ref{fig:ff-vs-redshift}. Even a $0.05hMpc^{-1}$ mode, which is short enough to be at risk of foreground contamination, maps to a far field distance of close to 800 meters.  For a global experiment the spectral evolution is set by the rate of change of the global temperature one wishes to be able to rule out.

The instrument (and by extension any drone calibration) must be designed to optimize performance on spatial scales out to 150m which places the far field at ranges between 1000 and 126km depending on instrument. Nearly all measurements will be operating decidedly inside the near field.

\section{Drones and Near Field Beam Mapping}
\label{sec:drones_and_mapping}

\subsection{Measuring the Pattern}
The as-built beam must be measured to demonstrate that it is free from the type of defects which lead to spectral structure at the modes we care about for cosmology, and at a low enough level such that lossy fitting or filtering is not required. In other words, the more convincingly instrumental artifacts are ruled out, the stronger evidence for any detections becomes.   

Several methods have been explored for capturing the pattern of a fixed antenna that can not be steered. Tradeoffs between astronomical objects, testing range, satellites, and more are described in \cite{Jacobs:2017}. Recent examples include solar and pulsar holography of the CHIME telescope \citep{Amiri:2022, Amiri:2024} and mapping of the MWA with the ORBCOMM satellite constellation\citep{Line:2018,Chokshi:2021}

With an unmanned aerial system (UAS) or ``drone''  the transmitter is under control of the experimenter subject to practical limitations like weather, legal rules, and mechanical reliability.  Difficulties can be overcome with engineering solutions but only if the cost matches the need. 

Drone measurements have been most frequently used to validate electromagnetic models or other measurements\citep{Acedo:2018,Zhang:2021}.  Drone validation of the CHIME beam has found broad agreement with the astronomically derived maps\cite{Tyndall:2025} as have measurements of the SKA low station\cite{Virone:2022}. In these measurements the investigators report broad satisfaction with the agreement between drone and simulation or other measurement. However, with no goal accuracy criteria it is difficult to judge the severity of quantitative disagreements.

For example, in drone measurements of SKA low antennas indicated a strong resonance due to antenna interactions \citep{paonessa:2023}. Using just a few tracks those authors found good agreement at some frequencies and angles but also differences of more than 100\%. Despite these differences the result was declared to be ``good agreement''. The differences between the model and the observation indicate that other factors, such as soil variation, are at work.

These divergences from model predictions raise questions about what a wider coverage across the sky hemisphere, would find. The horizon response is of particular interest as the response in that direction maps to the edge of the wedge and sets levels of mutual coupling.

\subsection{Short Review Drone mapping for 21cm Experiments}
This decade has seen significant progress towards drone-based antenna characterization. Here I make a brief review of some of this work with a focus on 21cm instruments and recent efforts towards mapping in the near field.

Drone-based radio calibration has been explored for application to wide-field 21~cm instruments. A drone-mounted calibration source was used to verify the accuracy of antenna response modeling for SKA-low stations \cite{2014IAWPL..13..169V} and then a second generation setup was then used as a phase calibrator \citep{Pupillo:2015}.  A first full mapping the beam of an antenna was  demonstrated by \citet{2015PASP..127.1131C}. Systems which recover high fidelity far fields have since been demonstrated \cite{Punzet:2022,Mauermayer:2022}. 

One common feature of many drone-based mapping systems is that the antenna under test operates at a relatively high frequency, here defined as above 1 GHz. At high frequencies a directional horn is of a size which can be lifted by a drone. At low frequencies (10-600 MHz) used for high redshift 21cm experiments, horns and other highly directive elements become prohibitively large and heavy for flight applications, motivating the use of electrically compact antennas, such as dipoles like those used by \citet{2014IAWPL..13..169V}. Fixing the transmitter antenna to the drone without a gimbal further reduces structural complexity and weight, but increases the need to understand the transmitter beam pattern since the orientation of the transmitting antenna will vary relative to the AUT.

In 2016 we tested a prototype system designed to make full hemispherical maps in the 21cm band\citep{Jacobs:2017}. A pair of calibration dipoles were mapped at 137MHz using a drone equipped with a short dipole and a CW transmitter. Simultaneously the difference between the two antennas was mapped with the ORBCOMM satellite array using the system developed by \citet{Neben:2016} for a comparison standard. Comparing these two maps the drone measurements agreed with the satellite but had smaller error bars at 2\% or better. This test map remains the largest and most accurate beam map yet made.

More recently, drone mapping in support of 21cm experiments have operated in the near field or explored near-field methods. These are discussed in the next section.

\subsection{Drones in the Near Field}

Reconstruction of the far field from near-field measurements is an area which has received much development over the past 30 years. See \citet{Breinbjerg:2016} for a review.  Theoretically, the far field can be inferred if both amplitude and phase of the E-field is sufficiently well sampled across a range of angles. There are also approximate methods if phase is not known. Since the far field is related to the near via a Fourier transform, measurement errors and non-uniform sampling result in non-local errors. Phase is challenging to measure directly.  For this reason, methods have been developed which use only amplitude \cite{Wang:2022}. These methods typically require measurement points made at multiple distances or with multiple probes and have reached a high level of technical development, particularly in the context of anechoic chambers but also with drones\cite{Garcia-Fernandez:2017}. 

Phase can be measured in absolute terms with synchronized clocks, in comparison to a reference antenna, differentially across frequency, or directly with a closed system such as in a network analyzer. 

The closest one can get to traditional near-field probing techniques is using a drone to scan a vector network analyzer probe\cite{Punzet:2022,Mauermayer:2022}. These methods require the VNA transmitter to be connected to the drone with a cable. This imposes a strict constraint, limiting operations to verticals scan in front of the antenna or within a relatively small volume.  Dense planar grid scans are needed to obtain fully hemispherical far field gain patterns per established 21cm requirements. Scanning a horizontal aperture tens to hundreds of meters wide with a cabled drone presents substantial practical difficulty.

Near-field drone measurements of larger horizontal apertures have been demonstrated on a SKA-low verification prototype by \cite{Ciorba:2019, Ciorba:2022}. Using a distant  antenna as a reference, the drone scanned across at a height of 10 wavelengths, a distance judged sufficiently far to avoid standing waves and reactive effects. This distance must be experimentally determined, but is usually a few wavelengths\cite{Anthony:2017}.  The measured phase agreed to within 3\% of the predicted value but amplitude could not be measured due to strong local interference. 

A phase calibrator reference antenna must itself be calibrated better than the desired gain error, or configured in such a way as to remove sources of phase variation.  The reference antenna will unavoidably have its own physical distortions on scales similar to those of interest in the antenna under test.  Such problems are not discussed much yet in the literature. 

Another novel way to measure in the near field is by actively phasing to the near field transmitter. In a test of this method using a LOFAR high band tile, the far-field reconstruction agreed with the model when the bore angle satisfied the small angle approximation but diverged at large angles\cite{Virone:2021}. This was potentially due to breakdown of the spatial approximation regime in which the near-field focusing method is valid\cite{Sherman:1962}. More importantly, near field focusing only removes phase curvature due to different path lengths from drone to array elements. Phasing does not remove path length differences due to passive interactions; these effects remain in the far field.

The differential frequency method ties every beam measured to a reference frequency which must be measured with other methods.  The differential position method requires more transmitters on the drone, adding size, weight, and complexity, or multiple flights.  The measured gradient must then be transformed into the far field using Fourier methods which are sensitive to missing data. The propagation of error would need to be carefully studied.

The difficulties and limitations of differential phase measurements and wired VNA systems motivate investigation of other ways of measuring absolute phase. One possibility is using a carrier coded with a known pattern. Use of psuedo random noise (PRN) coded signals are commonplace in Position Navigation and Timing systems such as global positioning as well as radar, for example the SIMONE meteor reflection system\cite{Chau:2021}. A prototype PRN beam mapping system is described by \citet{Bhopi:2023} which uses a software-defined radio to generate a PRN signal on the transmitter and correlates with a clone running on the receiver. This system has now been demonstrated, but  clock instability on time scales greater than 1s prevented recovery of absolute phase\citep{Bhopi:2026}.  

Another advantage of the PRN system is that the signal-to-noise of a matched filter is not biased by noise as in a total power measurement. Unintended radiation from the drone itself can add noise which varies according to position. This will add a systematic bias to a total power measurement, but averages down in correlation with the PRN reference. PRN correlation was demonstrated by \citet{Bhopi:2026} to significantly improve SNR allowing measurement further into the sidelobes of the beam even the correlation was only stable on short time scales.

One challenging requirement of a PRN system is the need for time synchronization between transmitter and receiver.  Timing stability of drone based calibrators were studied by \citet{Krishna:2024} who found that suitably portable GPS Disciplined Oscillators showed timing errors equivalent to more than 45$^\circ$ of phase, well beyond the usually accepted 22.5$^\circ$ and that stability degraded on timescales much longer than a minute.  Other sources of  jitter including receiver packet loss and temperature drift were not studied. These and other similar limitations would need to be addressed.

\section{Discussion of Drone Beam Mapping}
\label{sec:drone_prescription}
What beam mapping system is most likely to meet 21cm experiments operating below 200MHz?  Simplified to its most compact form the requirement is: Map amplitude and phase in near field sufficient to recover the far field pattern to the precision required by the experiment. This precision requirement depends on the net other aspects of the cosmology experiment such as sensitivity and uv sampling so it should be projected using a detailed forward model. For experiments that have been studied, the required beam amplitude accuracy is a few hundredths of a percent \cite{Jacobs:2017, Shaw:2015,Ewall-Wice:2017}.

Many methods for near-field measurement have been devised and here we have discussed a representative sample. Given that all have unique limitations and costs, a combination of methods is likely to be needed. 

The limitations of the PRN calibrator method and reference antenna methods are fairly independent.  The transmitted PRN signal can be observed with a reference antenna and correlated with the locally generated PRN just like the AUT. Indeed the PRN system offers a way to measure the beam pattern of the calibrator and conversely the phase reference system can be used to measure PRN clock stability. 

The phase reference antenna should be configured to minimize its own large scale phase structure. This could include being of higher quality construction and being arranged such that the geometric paths do not vary much during flight.  For example if the drone pattern can be kept small, so that the angular sweep through the reference antenna pattern is relatively small, than any phase variation might also be kept to a minimum. This is a subject for future study.

However it may well be that, as was found in anechoic chamber measurements, phase capture is brittle to small changes in cables and other difficult-to-manage subtleties making phaseless measurement techniques necessary after all (see \cite{Wang:2022}). In that case longer flight durations and more repeatable positioning will become necessary to support multiple layers and intersecting measurement grids.

\section{Summary and Future Work}
Detection of the 21\,cm cosmological signal requires new levels of precision, particularly along the
spectral dimension at scales between 1 and 20MHz across the 50 to 200MHz cosmology band. This is demonstrably true for interferometers and true for the global signal if the goal is to rule out the type of sharp-edged trough signal reported by EDGES.   This spectral scale corresponds to physical scales up to 100m.

Global instruments have demonstrated several approaches to controlling large scale structure: EDGES with a very large ground plane and by covering the nearby hut with absorber, SARAS3 floating on a lake, or the newly proposed EIGSEP hanging over a canyon. Similarly, large arrays must also adapt to minimize large scale structure, for example CHORD, operating as low as 300MHz, has adopted an extremely deep focus specifically to minimize mutual coupling\cite{Vanderlinde:2019}.  Regardless of the design approach, the performance must be checked with independent measurements. Verification with a drone must be done in the near field, necessitating a more challenging measurement such as phase capture. Work to date on drone-based phase measurement has shown  promise and difficulties ahead.

This analysis made several approximations in the interest of coming to an estimate of the cosmological requirements. The theory signal was modeled with an emulator based on a parametric model that didn't include exotic ingredients or more detailed physical simulations. Brighter signals will have higher maximum delays mapping to larger physical scales in the instrument. When converting to instrumental units the baseline was kept short compared to the line of sight k mode. A real instrument will also have longer baselines which sample sections of the power spectrum in slices at higher $k_\perp$ They were neglected here because the 21cm amplitude signal decreases with $k$ and also to allow us to assume an identity window function mapping frequency to delay.  Similarly, the sensitivity estimate, needed to set an upper bound on delays which are theoretically relevant, could be done with more accuracy or for a range of instruments. Decrease in the noise would have the effect of increasing the maximum delays and size scales that would have to be characterized. All of these variables have a relatively small effect on the question of relevant size scales that determine the far field.

\section*{Acknowledgments}
The author was supported by the NSF under grant \#2144995 
\bibliographystyle{plainnat}
\bibliography{bibtex/bib/ECHO2024}

\end{document}